\documentclass[twocolumn]{aastex631}

\usepackage{amsmath}
\usepackage{booktabs}
\usepackage{ulem}
\shorttitle{st}
\shortauthors{Du \& Fan}
\begin{document}
\title{A Delayed Multi-channel Progenitor for Apparently Nonrepeating Fast Radio Bursts}
\author[0000-0002-4375-3737]{Zhao-Wei Du}
\affiliation{School of Physics Science and Technology, Wuhan University, No.299 Bayi Road, Wuhan, Hubei, China; xilong.fan@whu.edu.cn}

\author[0000-0002-8174-0128]{Xi-Long Fan} 
\affiliation{School of Physics Science and Technology, Wuhan University, No.299 Bayi Road, Wuhan, Hubei, China; xilong.fan@whu.edu.cn}

\author[0009-0006-5136-2011]{Yi-Xiao Li} 
\affiliation{School of Physics Science and Technology, Wuhan University, No.299 Bayi Road, Wuhan, Hubei, China; xilong.fan@whu.edu.cn}

% \author[0000-0001-6396-9386]{HouJun L\"{u}} 
% \affiliation{Guangxi Key Laboratory for Relativistic Astrophysics, Department of Physics, Guangxi University, Nanning 530004, China;}

\begin{abstract}
Fast radio bursts (FRBs) are millisecond-duration radio flashes of unknown origin, observationally classified into repeating and apparently nonrepeating (one-off) populations. In this work, we use a statistical population approach  to investigate the redshift evolution of one-off FRBs. We compare a pure star formation history (SFH) tracing model, phenomenological delayed models, physically motivated delayed models that correspond to binary neutron star related and neutron star age-window channels, and mixture models which is obtained when two physically motivated models are normalized separately and then combined as a weighted mixture.  The samples in CHIME/FRB Catalog do not support an intrinsic event rate density that directly follows the SFH.  The preferred model is mixture model corresponds to an effective mean delay time of $\bar{\tau}=1.426^{+0.032}_{-0.035}~\mathrm{Gyr}$. These results suggest that the current data may naturally explained by delayed, possibly multi-channel progenitor evolution for the one-off FRBs.
\end{abstract}

\section{Introduction}\label{sec:intro}
FRBs are millisecond-duration radio transients characterized by large dispersion measures, high brightness temperatures, and predominantly extragalactic or cosmological distances \citep{2013Sci...341...53T, 2019A&ARv..27....4P, 2023RvMP...95c5005Z}. The Lorimer burst marked the beginning of FRB astronomy as the first identified one-off FRB, whereas FRB~121102, the first confirmed repeating FRB, demonstrated that at least some FRB engines can remain active after producing an observable burst \citep{2007Sci...318..777L, 2016Natur.531..202S}. The detection of both repeating and apparently nonrepeating sources further suggests that FRBs are not a phenomenologically uniform class.

In recent years, wide field and high sensitivity radio telescopes, including CHIME and FAST, have transformed FRB studies from individual event discovery to statistical population analysis. The first CHIME/FRB catalog reported 536 bursts detected within roughly one year of operation, providing the first large FRB sample observed with a single instrument and relatively uniform selection effects \citep{2021ApJS..257...59C}. The second CHIME/FRB catalog further enlarged the sample to 4539 bursts from 3641 unique sources, including 981 bursts from 83 known repeating sources \citep{2026ApJS..283...34C}. High-sensitivity FAST observations have substantially expanded burst statistics for active repeating FRBs, allowing their energy distributions, polarization behaviour, and local magneto-ionic environments to be studied in much greater detail \citep{2021Natur.598..267L, 2022Natur.606..873N, 2022Natur.609..685X}. More broadly, the rapid expansion of FRB surveys, follow-up campaigns, and localization programs has greatly advanced our understanding of the FRB population \citep{2014ApJ...790..101S, 2019Sci...365..565B, 2024ApJ...967...29L}. These observational advances have enabled systematic studies of properties of FRBs. Such large and increasingly homogeneous samples are expected to play a central role in clarifying the still mysterious physical origin of FRBs. Establishing how the observed FRB population is connected to these possible source channels remains one of the central problems in FRB astrophysics.

The observational distinction between repeating and one-off FRBs is therefore an important starting point for population studies. Repeating FRBs demonstrate that at least some engines can remain active after producing an observable burst, and several well-studied repeaters show diverse observational properties \citep{2016Natur.531..202S, 2017Natur.541...58C, 2019ApJ...885L..24C, 2020Natur.587...54C, 2021ApJ...923....1P, 2022Natur.609..685X}. By contrast, most FRBs detected in wide-field surveys have not been observed to repeat. Large sample comparisons have further shown that bursts from repeating sources tend to have longer temporal widths and narrower emission bandwidths than apparent nonrepeaters  \citep{2021ApJ...923....1P, 2025ApJ...992..206C}. Although the lack of detected repetition is an observational classification rather than definitive evidence for a distinct physical class, the statistical differences between repeaters and one off FRBs suggest that the two groups may either trace different progenitor channels or represent different activity states within a broader underlying source population. The Galactic event FRB~200428, associated with the magnetar SGR~1935+2154, provided direct evidence that neutron star magnetospheric activity can produce FRB-like radio bursts \citep{2020Natur.587...54C, 2020Natur.587...59B, 2020ApJ...898L..29M, 2021NatAs...5..372R, 2021NatAs...5..378L}. Motivated by these observational clues, the origin of one off FRBs remains a particularly important problem: they dominate current survey samples, but their physical interpretation is still uncertain.

Several classes of progenitor models have been proposed for one off FRBs. Magnetar related models attribute FRB emission to magnetospheric activity, magnetic reconnection, or relativistic outflows from young or active neutron stars, and are supported observationally by the Galactic FRB-like event FRB~200428 from SGR~1935+2154 \citep{2020Natur.587...54C, 2020Natur.587...59B, 2020ApJ...898L..29M}. Collapse models, such as the 'blitzar' scenario, associate a one-off burst with the collapse of a supramassive neutron star into a black hole, producing a short electromagnetic pulse as the magnetosphere is expelled \citep{2014A&A...562A.137F, 2018ApJ...864..117M}. Compact binary models instead connect one off FRBs with neutron star mergers, neutron star--black hole mergers, or related magnetospheric interactions shortly before or after coalescence \citep{2001MNRAS.322..695H, 2013PASJ...65L..12T, 2015ApJ...814L..20M, 2016ApJ...822L...7W, 2019ApJ...873L...9Z}. Beyond the main classes discussed above, several more speculative scenarios have also been proposed, typically involving less conventional compact object systems or more exotic physical mechanisms \citep{2013ApJ...776L..39K, 2017JHEAp..13...22R, 2014PhRvD..90l7503B}. These models differ not only in their emission mechanisms, but also in their expected evolution: some channels may approximately follow recent star formation, whereas others naturally involve long delay times. Testing these possibilities statistically therefore requires information on the distance or redshift distribution of one off FRBs.

A number of statistical studies have therefore attempted to connect the observed FRB distributions with their underlying event-rate evolution \citep{2020MNRAS.494..665L, 2021MNRAS.501..157Z, 2022ApJ...924L..14Z, 2023ChPhC..47h5105T, 2024ApJ...969..123L}. In particular, \citet{2024ApJ...969..123L} used a statistical framework to compare FRB population models whose event rate histories either trace the SFH or are delayed with respect to it. Such studies showed that the observed FRB distributions contain useful information about source evolution. At the same time, they also made clear that the interpretation depends on the adopted sample, the assumed relation between extragalactic dispersion measure and redshift, the burst energy distribution, and the detection efficiency model. These conclusions are not yet fully settled across different analyses. For example, \citet{2021MNRAS.501..157Z} tested SFH-tracing and compact binary merger like redshift distributions using Parkes and ASKAP samples and found that the available data could not reject any of the considered redshift distribution models. In contrast, \citet{2022ApJ...924L..14Z} argued that the CHIME FRB population does not simply track the SFH and is better described by a delayed or hybrid evolutionary model. More recently, \citet{2025ApJ...986..100G} used the CHIME/FRB Baseband Catalog 1 to calibrate the joint fluence--dispersion-measure distribution and found that the redshift distribution of an one off FRB samples are more naturally explained by a mixed evolutionary scenario, in which a young component related to star formation is supplemented by an older or delayed channel. \citet{2023ApJ...944..105S} inferred the energy and distance distributions of CHIME FRBs, but found that the evolution of the FRB population remained poorly constrained.

The release of the second CHIME/FRB catalog provides a new opportunity to revisit these questions with improved statistics and more uniform data products. Catalog 2 greatly enlarges the CHIME sample and provides consistently processed measurements of burst properties, including dispersion measure, fluence-related quantities, temporal width, scattering time, localization, exposure, and sensitivity information \citep{2026ApJS..283...34C}. This is particularly useful for studying one-off FRBs, for which individual host identifications are generally unavailable but collective distributional properties can be measured with high statistical power. At the same time, the redshift inference of such sources remains sensitive to the adopted relation between extragalactic dispersion measure and redshift, motivating the use of an updated relation in the present analysis \citep{2026arXiv260318487C}. We therefore use Catalog 2 as a statistical population sample to examine whether the observed fluence, energy, and inferred redshift distributions of one-off FRBs are more naturally reproduced by an SFH-tracing model, delayed phenomenological models, or physically motivated mixed-evolution models.

In this work, we use the CHIME/FRB Catalog 2 Gold Sample to study the population properties of one-off FRBs. We statistically infer their redshift distribution and jointly model their fluence, isotropic energy, and redshift distributions. We compare several prescriptions for the intrinsic FRB event rate density, including an SFH model, phenomenological delayed models, and physically motivated models. The paper is organized as follows. In Section~\ref{sec:meth}, we describe the $\mathrm{DM}_{\rm E}$--$z$ relation, the burst-energy distribution, the detection-efficiency model, and the event-rate prescriptions. In Section~\ref{sec:data}, we introduce the CHIME/FRB Catalog 2 Gold Sample, describe the selection criteria, present the fitting results, and compare the statistical performance of the different population models. In Section~\ref{sec:conclusion}, we discuss the physical implications and limitations of the inferred evolutionary behavior.

\section{METHODOLOGY} \label{sec:meth}

\subsection{$\text{DM}_{\rm E}-z$ relation}
A detailed investigation of the FRB population relies on redshift information for individual bursts. However, only a limited subset of FRBs currently have confirmed redshift measurements, primarily because most bursts lack sufficiently precise localization for robust host-galaxy identification. As a result, redshift information for the majority of bursts must be inferred statistically rather than determined directly. To address this limitation, some papers introduce a probabilistic relation that connects the dispersion measure to redshift. The  conditional probability density of the extragalactic dispersion $\mathrm{DM}_E$ at a given redshift $z$  can be written as \citep{2020Natur.581..391M}
\begin{equation}
\begin{aligned}
p_E(\mathrm{DM}_E|z) &= \int_0^{(1+z)\mathrm{DM}_E} p_{\mathrm{host}}(\mathrm{DM}_{\mathrm{host}}|\mu,\sigma_{\mathrm{host}})\\[1ex]
&\quad \times p_{\mathrm{IGM}}\!\left(\mathrm{DM}_E-\frac{\mathrm{DM}_{\mathrm{host}}}{1+z}\,\middle|\,F,z\right) \,d\mathrm{DM}_{\mathrm{host}}.
\end{aligned}
\end{equation}
The integral marginalizes over the unknown host contribution, represented by $p_{\mathrm{host}}$, while $p_{\mathrm{IGM}}$ describes the corresponding contribution from the intergalactic medium, whitch  can be constrained using well-localized FRBs \citep{2023ChPhC..47h5105T, 2026arXiv260318487C}. $\mathrm{DM}_E$ is obtained by subtracting the Milky Way interstellar medium and halo contributions from the observed dispersion measure, and can be expressed as \citep{2014ApJ...783L..35D, 2014ApJ...788..189G, 2020Natur.581..391M}
\begin{equation}
\begin{aligned}
\mathrm{DM}_{\mathrm{E}} &\equiv \mathrm{DM}_{\mathrm{obs}} - \mathrm{DM}_{\mathrm{MW,ISM}} - \mathrm{DM}_{\mathrm{MW,halo}} \\
&= \mathrm{DM}_{\mathrm{IGM}} + \frac{\mathrm{DM}_{\mathrm{host}}}{1+z},
\end{aligned}
\end{equation}
where $\mathrm{DM}_{\mathrm{IGM}}$ and $\mathrm{DM}_{\mathrm{host}}$ denote the contributions from the intergalactic medium and the host galaxy, respectively. The Milky Way interstellar-medium contribution, $\mathrm{DM}_{\mathrm{MW,ISM}}$, is estimated using the NE2001 and YMW16 electron density models \citep{2002astro.ph..7156C, 2017ApJ...835...29Y}, while the Milky Way halo contribution, $\mathrm{DM}_{\mathrm{MW,halo}}$, is fixed at $50~\mathrm{pc\,cm^{-3}}$ \citep{2019MNRAS.485..648P, 2020Natur.581..391M}.

Given that the properties of FRB host galaxies are still poorly constrained, we adopt a phenomenological prescription for the distribution of $\mathrm{DM}_{\mathrm{host}}$, written as \citep{2020Natur.581..391M, 2020ApJ...900..170Z}
\begin{equation}
    \begin{aligned}
p_{\text{host}}(\text{DM}_{\text{host}}|\mu, \sigma_{\text{host}}) &= \frac{1}{\sqrt{2\pi}\,\text{DM}_{\text{host}}\sigma_{\text{host}}} \\
&\quad \times \exp\left[-\frac{(\ln\text{DM}_{\text{host}} - \mu)^2}{2\sigma_{\text{host}}^2}\right],
\end{aligned}
\end{equation}
whereas $\mu$ and $\sigma_{\mathrm{host}}$ are mean and standard deviation of $\ln \mathrm{DM}_\mathrm{host}$, respectively.

The probability distribution of the intergalactic-medium contribution can be calibrated from numerical simulations and is well described by a quasi-Gaussian form \citep{2020Natur.581..391M, 2021ApJ...906...49Z},
\begin{equation}
    p_{\rm IGM}(\Delta)=A \Delta^{-3}\exp\left[-\frac{(\Delta^{-3}-C_0)^2}{18\sigma_{\rm IGM}^2}\right],\quad \Delta>0,
\end{equation}
where $\Delta \equiv \mathrm{DM}_{\mathrm{IGM}}/\langle\mathrm{DM}_{\mathrm{IGM}}(z)\rangle$, $\sigma_{\rm IGM}=Fz^{-1/2}$ with $F$ being a free parameter, $A$ is the normalization constant, and $C_0$ is chosen such that the mean of $\Delta$ remains unity. The mean IGM dispersion measure is obtained by integrating the cosmological free-electron column density along the line of sight. It is given by \citep{2014ApJ...783L..35D, 2014ApJ...788..189G, 2014PhRvD..89j7303Z, 2020Natur.581..391M}
\begin{equation}
    \langle\mathrm{DM}_{\mathrm{IGM}}(z)\rangle = \frac{3cH_0\Omega_b f_{\rm IGM}f_e}{8\pi G m_p}\int_0^z \frac{1+z'}{\sqrt{\Omega_m(1+z')^3+\Omega_\Lambda}}\,dz',
\end{equation}
where $c$ is the speed of light, $m_p$ is the proton mass, $G$ is the Newtonian gravitational constant, $f_e=7/8$ is the electron fraction, and $f_{\rm IGM}=0.84$ is the baryon mass fraction in the IGM. The cosmological parameters $H_0$, $\Omega_b$, and $\Omega_\Lambda$ are fixed to the Planck 2018 values. 

The intrinsic distributions predicted by a population model are not expected to coincide directly with the observed ones, because the observed FRB sample is shaped by instrumental sensitivity and selection effects. Therefore, in order to make a meaningful comparison between the theoretical model and the CHIME data, the detection process must be taken into account. After incorporating the selection effect, the observed distributions of fluence, energy, and redshift can be written as \citep{2024ApJ...962...73L}
\begin{equation}
    \begin{aligned}
p_{\rm det}(F_\nu) &\propto \frac{d}{dF_\nu} \iint\limits_{F_\nu(E,z)\leq F_\nu} p(z)\,p(E)\,\eta_{\rm det}(F_\nu(E,z))\,dE\,dz, \\
p_{\rm det}(E) &\propto \int_{z_{\rm min}}^{z_{\rm max}} p(z)\,p(E)\,\eta_{\rm det}(F_\nu(E,z))\,dz, \\
p_{\rm det}(z) &\propto \int_{E_{\rm min}}^{E_{\rm max}} p(z)\,p(E)\,\eta_{\rm det}(F_\nu(E,z))\,dE.
\end{aligned}
\end{equation}
$p_{\rm det}(F_\nu)$, $p_{\rm det}(E)$, and $p_{\rm det}(z)$ denote the predicted observed distributions of fluence, energy, and redshift after considering the selection effects. $p(z)$ denotes the intrinsic redshift distribution of the FRB population, the explicit form of which will be discussed in the next subsection, while $p(E)$ denotes the intrinsic burst-energy distribution. As suggested by previous work, the burst-energy distribution can be characterized by a power-law form with an exponential suppression at high energies \citep{2018MNRAS.481.2320L, 2020MNRAS.494..665L, 2021MNRAS.501..157Z},
\begin{equation}
    p(E) \propto \left(\frac{E}{E_c}\right)^{-\alpha} \exp\left(-\frac{E}{E_c}\right),
\end{equation}
The parameters $\alpha$ and $E_c$ are left free and constrained by the fit. $\eta_{\rm det}(F_\nu(E,z))$ is modeled by \cite{2022ApJ...924L..14Z} They assumes a gray zone between a minimum threshold fluence, $F_{\nu,\mathrm{th}}^{\min}$, and a maximum threshold fluence, $F_{\nu,\mathrm{th}}^{\max}$, within which the detection probability rises gradually from 0 to 1. The expression can thus be written as 
\begin{equation}
\eta_{\mathrm{det}}(F_\nu)=
\begin{cases}
0, & \log F_\nu \le \log F_{\nu,\mathrm{th}}^{\min},\\
\left(\dfrac{\log F_\nu-\log F_{\nu,\mathrm{th}}^{\min}}
{\log F_{\nu,\mathrm{th}}^{\max}-\log F_{\nu,\mathrm{th}}^{\min}}\right)^n,
& \mathrm{otherwise} ,\\
1, & \log F_\nu \ge \log F_{\nu,\mathrm{th}}^{\max},
\end{cases}
\end{equation}
where $n$ and $\log F_{\nu,\mathrm{th}}^{\max}$ are treated as free parameters, while $\log F_{\nu,\mathrm{th}}^{\min}=-0.5$ is fixed following \citet{2022ApJ...924L..14Z}. The observed specific fluence corresponding to a burst with energy $E$ at redshift $z$ can be written as \citep{2018MNRAS.480.4211M, 2022MNRAS.509.4775J}
\begin{equation}
    F_{\nu}(E, z) = \frac{(1+z)^{2+\beta}}{4\pi d_{L}^{2}\Delta\nu}\,E.
\end{equation}
In this work, we fix $\beta=-1.5$ \citep{2019ApJ...872L..19M}, $d_L$ denotes the luminosity distance, and $\Delta\nu=400~\mathrm{MHz}$ is the observing bandwidth \citep{2021ApJS..257...59C, 2026ApJS..283...34C}.

\subsection{the intrinsic redshift distribution of FRBs }  
The observed redshift distribution of FRBs is related to the intrinsic event rate density $R(z)$ through the conversion from comving frame time to observer time and the differential comoving volume element $dV_c/dz$. The realtion can be experssed as
\begin{equation}
    p(z)\propto \frac{R(z)}{1+z} \frac{dV_c}{dz}.
\end{equation}
In the flat $\Lambda$CDM cosmology, the differential comoving volume element can be written as
\begin{equation}
    \frac{dV_c}{dz} = \frac{4\pi c d_{L}^{2}}{H_{0}(1+z)^2\sqrt{\Omega_{m}(1+z)^{3}+\Omega_{\Lambda}}}.
\end{equation}
Therefore, the main differences among the models lie in the form of $R(z)$. In this paper, we consider two classes of models: a phenomenological models following \citet{2024ApJ...969..123L}, and a physically motivated models. 

We first introduce the phenomenological models. Because the progenitor scenario of FRBs remains uncertain, two broad possibilities are commonly considered. If FRBs are produced promptly in core collapse events or by newly born neutron stars, their intrinsic event-rate density is expected to trace the SFH. By contrast, if the FRB engine requires a substantial evolutionary timescale before becoming active, as in scenarios involving binary neutron star (BNS) mergers or the newborn magnetar from merger remnants, the intrinsic event rate density should be described by the convolution of the SFH with a delay time distribution (DTD). Therefore, in the phenomenological framework considered here, the intrinsic event rate density either directly follows the SFH or is modeled as the SFH convolved with a DTD. In this work, we consider three commonly used forms of the DTD: the Gaussian model, the log-normal model, and the power-law model \citep{2021MNRAS.501..157Z, 2024ApJ...969..123L}.

\begin{enumerate}
    \item Gaussian delay model:
    \begin{equation}
        f(\tau)=\frac{1}{\sigma_{\rm G}\sqrt{2\pi}}
        \exp\left[-\frac{(\tau-\tau_{\rm G})^2}{2\sigma_{\rm G}^2}\right].
    \end{equation}

    \item Log-normal delay model:
    \begin{equation}
        f(\tau)=\frac{1}{\tau\sigma_{\rm log}\sqrt{2\pi}}
        \exp\left[-\frac{(\ln\tau-\ln\tau_{\rm log})^2}{2\sigma_{\rm log}^2}\right].
    \end{equation}

    \item Power-law delay model:
    \begin{equation}
        f(\tau)=
        \left(
        \frac{1-\alpha_\tau}
        {\tau_{\max}^{1-\alpha_\tau}-\tau_{\min}^{1-\alpha_\tau}}
        \right)\tau^{-\alpha_\tau}.
    \end{equation}
\end{enumerate}
Following \citet{2024ApJ...969..123L}, the minimum delay time is fixed at $\tau_{\min}=20~\mathrm{Myr}$, while the maximum delay time is taken to be $\tau_{\max}=1/H_0$.

 The experssion of $R(z)$ of phenomenological model is 
\begin{equation}
    R(z) \propto \int_{z}^{\infty} \mathrm{SFH}(z') \, f\!\left[t(z)-t(z')\right] \frac{dt}{dz'} \, dz'.\label{Rz}
\end{equation}
Here, $\mathrm{SFH}(z)$ is modeled as \citep{2017ApJ...840...39M}
\begin{equation}
    \mathrm{SFH}(z)= \frac{(1+z)^{2.6}}{1 + \left(\dfrac{1+z}{3.2}\right)^{6.2}}.
\end{equation}
$t(z)$ is cosmic time at redshift $z$, its experssion is 
\begin{equation}
    t(z) = \int_{z}^{\infty} \frac{1}{H_{0}(1+z^\prime)\sqrt{\Omega_{m}(1+z^\prime)^{3}+\Omega_{\Lambda}}} dz^\prime, 
\end{equation}
and the Eq. (\ref{Rz}) describes the contribution of progenitors formed at earlier epochs and observed as FRBs after a delay time $\tau=t(z)-t(z')$.

The progenitor(s) of FRBs remain unclear. Recent studies have suggested that the FRB population may not be described by a single channel, but may instead involve multiple formation pathways and/or delayed evolutionary timescales \citep{2022ApJ...939...27C, 2023ApJ...954...80G, 2023ApJ...944..105S, 2024ApJ...969..123L,2025ApJ...986..100G, 2025A&A...698A.127M}. Motivated by these developments, we consider three physically motivated models in this work. 

The first physically motivated scenario assumes that FRBs are associated either with binary neutron star merger events themselves or with young magnetars born in their remnants. Since the known Galactic FRB source is associated with a very young magnetar \citep{2016MNRAS.457.3448I, 2020Natur.587...54C}, it is reasonable to assume that, if FRB-producing magnetars can also be formed through BNS mergers, their active phase may occur sufficiently soon after birth that the corresponding FRB event rate density still approximately tracks the BNS merger rate. Their intrinsic event density is
\begin{equation}
\begin{aligned}
     R(z) \propto \int_{z}^{\infty} \tau^{-\alpha_{\tau,\rm BNS}}&\mathrm{SFH}(z^\prime)\left|\frac{dt}{dz^\prime}\right|dz^\prime,\\
     &t_{\min,\rm BNS} \le \tau \le t_{\max,\rm BNS}.
\end{aligned}
\end{equation}
The second physically motivated scenario assumes that FRBs are produced by neutron stars within a particular age range; that is, the FRB-producing neutron stars are assumed to lie within an age window $[A_{\min},\,A_{\max}]$. To implement this scenario, we first calculate the neutron star birth rate density from the SFH, weighted by the progenitor mass function. We then integrate this birth-rate density over the corresponding birth time interval to obtain the number density of neutron stars whose ages fall within the adopted age window. This can be written as

\begin{equation}
R(z)\propto
\int_{t(z)-A_{\max}}^{t(z)-A_{\min}}
\mathrm{SFH}(z^\prime(t^\prime))\,dt^\prime.
\end{equation}

The third physically motivated scenario is a hybrid model that combines the above two channels. In this case, we first construct the redshift distributions associated with the BNS related channel and the age window neutron star channel, and normalize each of them separately. The final intrinsic redshift distribution is then written as a weighted sum of the two normalized components,
\begin{equation}
p_{\rm mix}(z)=w_{\rm BNS}\,p_{\rm BNS}(z)+\left(1-w_{\rm BNS}\right)p_{\rm NS~age}(z).
\end{equation}
Here, $w_{\rm BNS}\in[0,1]$ is a free mixture parameter that controls the relative contribution of the BNS related channel, while $1-w_{\rm BNS}$ gives the corresponding weight of the age window neutron star channel. Since the two components are individually normalized before combination, $w_{\rm BNS}$ should be interpreted as a mixture weight in redshift distribution space, rather than as a strict astrophysical branching fraction or physical event rate ratio. 

From this perspective, the first two physically motivated models may be viewed as physically interpretable restrictions of the phenomenological power-law delay and SFH-based descriptions. The BNS-related model preserves the form of an SFH convolved power-law delay model, while assigning it a specific progenitor interpretation in terms of BNS mergers and their remnants. The neutron star age window model, on the other hand, replaces a generic SFH tracing population with a restricted subset of neutron stars selected by age. In this sense, these models do not simply add phenomenological flexibility, but instead impose physically motivated constraints on the underlying redshift evolution. 

\subsection{the populaiton distribution}

After obtaining the model-predicted observed distributions of fluence, energy, and redshift, we constrain the model parameters by comparing them with the CHIME/FRB sample. Rather than fitting the differential probability density functions directly, we use the corresponding cumulative number counts above a given threshold, which avoids the arbitrariness associated with binning. For a generic observable $Q\in\{F_\nu, E, z\}$, the model-predicted cumulative count is written as
\begin{equation}
N_{\rm det}(>Q)=A\int_{Q}^{Q_{\max}} p_{\rm det}(Q')\,dQ',
\end{equation}
where $A$ is a normalization constant chosen such that the total number of model events matches the sample size. The predicted cumulative counts are then compared with the observed cumulative counts $N(>Q_i)$, defined as the number of FRBs with values larger than the $i$-th data point $Q_i$. For each observable, we define a chi-square statistic as
\begin{equation}
\chi_Q^2=\sum_i \frac{\left[N_{\rm det}(>Q_i)-N(>Q_i)\right]^2}{\sigma_i^2},
\end{equation}
with the uncertainty taken to be $\sigma_i=\sqrt{N(>Q_i)}$, following Poisson counting statistics. The total goodness-of-fit is then constructed by summing the contributions from fluence, energy, and redshift,
\begin{equation}
\chi_{\rm total}^2=\chi_{F_\nu}^2+\chi_E^2+\chi_z^2.
\end{equation}
Assuming Gaussian errors in this effective $\chi^2$ sense, the likelihood of the model parameters $\boldsymbol{\theta}$ is taken to be
\begin{equation}
\mathcal{L}(\mathrm{FRBs}\mid\boldsymbol{\theta})\propto
\exp\!\left(-\frac{1}{2}\chi_{\rm total}^2\right),
\end{equation}
where $\boldsymbol{\theta}$ denotes the full set of free parameters in the redshift model, burst-energy distribution, and detection-efficiency function. The posterior distribution is then obtained through Bayes' theorem,
\begin{equation}
P(\boldsymbol{\theta}\mid \mathrm{FRBs})\propto
\mathcal{L}(\mathrm{FRBs}\mid\boldsymbol{\theta})\,P_0(\boldsymbol{\theta}),
\end{equation}
where $P_0(\boldsymbol{\theta})$ is the prior distribution of the model parameters. For all models, the common free parameters are $\alpha, E_c, \log F_{\nu,\mathrm{th}}^{\max}$ and $n$. For the phenomenological models, the Gaussian, log-normal, and power-law delay models include $\tau_{\rm G},\sigma_{\rm G}$, $\tau_{\log},\sigma_{\log}$, and $\alpha_\tau$, respectively. For the physically motivated models, the BNS-related model is described by $\tau_{\rm min,BNS},\alpha_{\tau,\rm BNS}$, the neutron star age-window model by $A_{\min},A_{\max}$, and the mixture model uses the parameters of both components with one additional mixture weight $w_{\rm BNS}$.

\section{Data and Results} \label{sec:data}

\subsection{the gold samples of FRBs}
In this work, we use the second CHIME/FRB catalog as the parent sample for constraining the model parameters \citep{2026ApJS..283...34C}. Unlike the first CHIME/FRB catalog \citep{2021ApJS..257...59C}, which reported 536 bursts in total, including 474 one-off bursts and 62 bursts from 18 repeating sources, the second catalog provides a much larger and more uniformly processed data set for population studies. It contains 4539 bursts detected between 2018 July 25 and 2023 September 15, originating from 3641 unique sources, including 981 bursts from 83 known repeating sources. Importantly, Catalog 2 fully includes all events from Catalog 1, but reprocesses them together with the newly detected bursts within improved analysis framework. For the analysis below, we restrict the sample to the one-off FRBs, since repeaters and one off FRBs may have different physical origins.

To improve the robustness of our analysis, we adopt the Gold Sample constructed from the second CHIME/FRB catalog, which contains 1973 events after applying a uniform set of quality selection criteria. Specifically, we retain only bursts with a reliable extragalactic dispersion measure estimate, $\mathrm{DM}_E>100~\mathrm{pc\,cm^{-3}}$. We further require a high signal-to-noise ratio ($\mathrm{SNR}>12$), a dominant extragalactic DM contribution ($\mathrm{DM}_{\mathrm{obs}}>1.5\max(\mathrm{DM}_{\mathrm{NE2001}},\mathrm{DM}_{\mathrm{YMW16}})$), no sidelobe contamination, no catalog exclusion flag, and a small scattering time ($\tau_{\mathrm{scat}}<0.01~\mathrm{s}$). In addition, when applying the updated $\mathrm{DM}_{\rm E}$-$z$ relation of \citet{2026arXiv260318487C}, we exclude sources with inferred redshifts $z>3$ in order to avoid extrapolating the adopted relation beyond the range where it is expected to remain reliable. The resulting Gold Sample therefore provides a cleaner and more homogeneous set of one off FRBs for the population analysis presented below.

\subsection{results}
\begin{figure}
\centering
\includegraphics[width=0.95\textwidth]{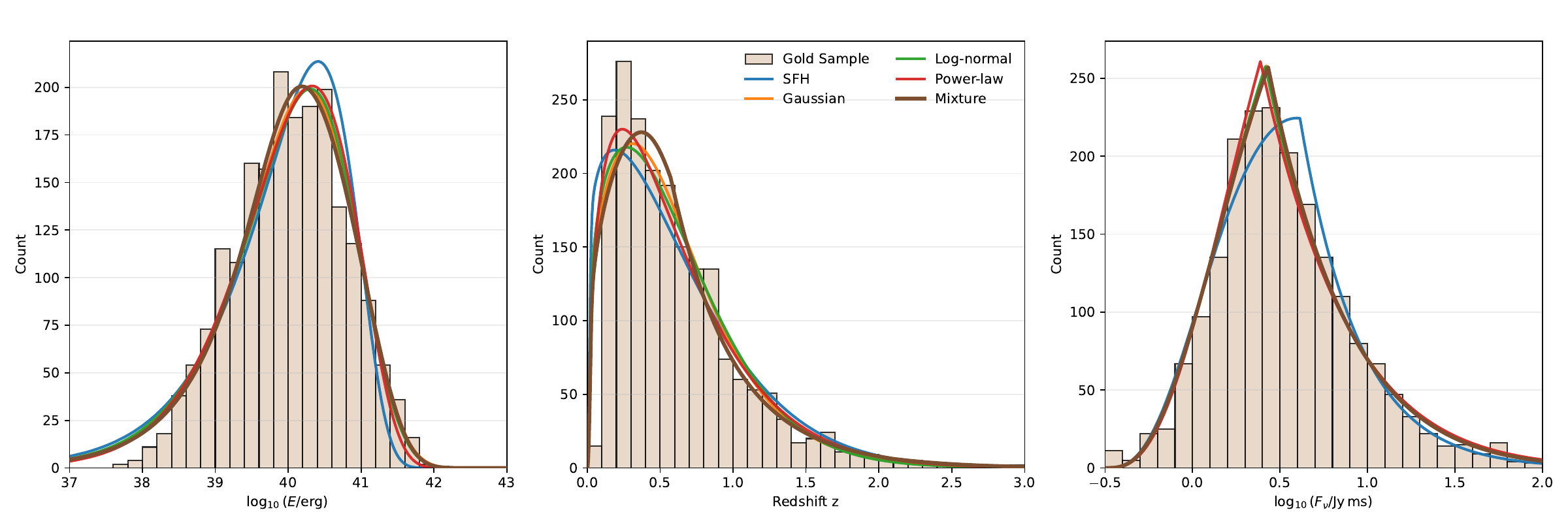}
\caption{
Comparison between the Gold Sample distributions and the model predictions for the phenomenological and mixture models. 
The three panels show the distributions of $\log E$, redshift $z$, and $\log F_{\nu}$, respectively. 
The pure SFH model shows the poorest agreement with the data, whereas the delayed models and the mixture model provide substantially improved descriptions.
}
\label{fig:mixture_overlay}
\end{figure}

\begin{figure*}[htbp]
\centering
\gridline{
    \fig{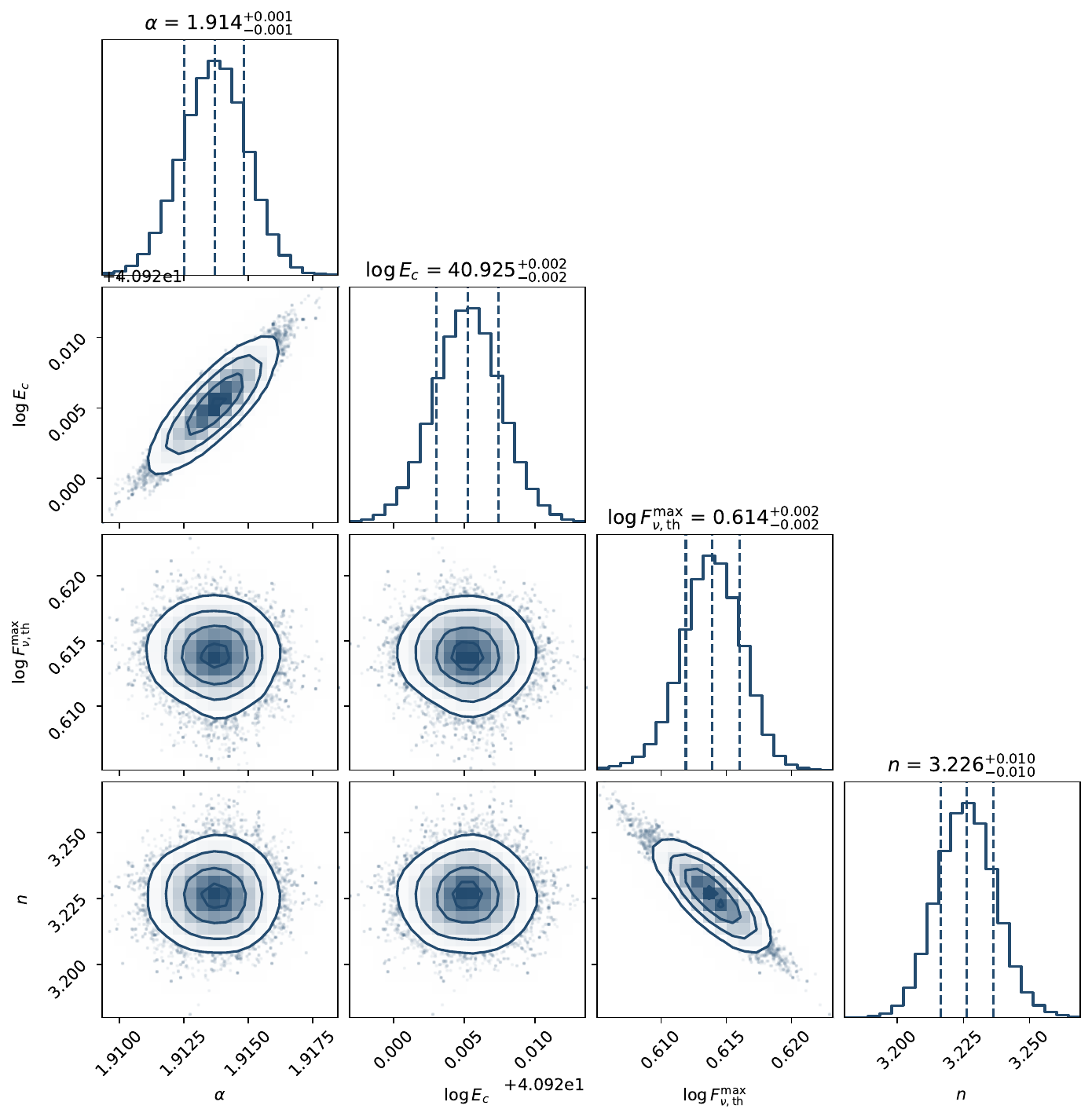}{0.48\textwidth}{}
    \fig{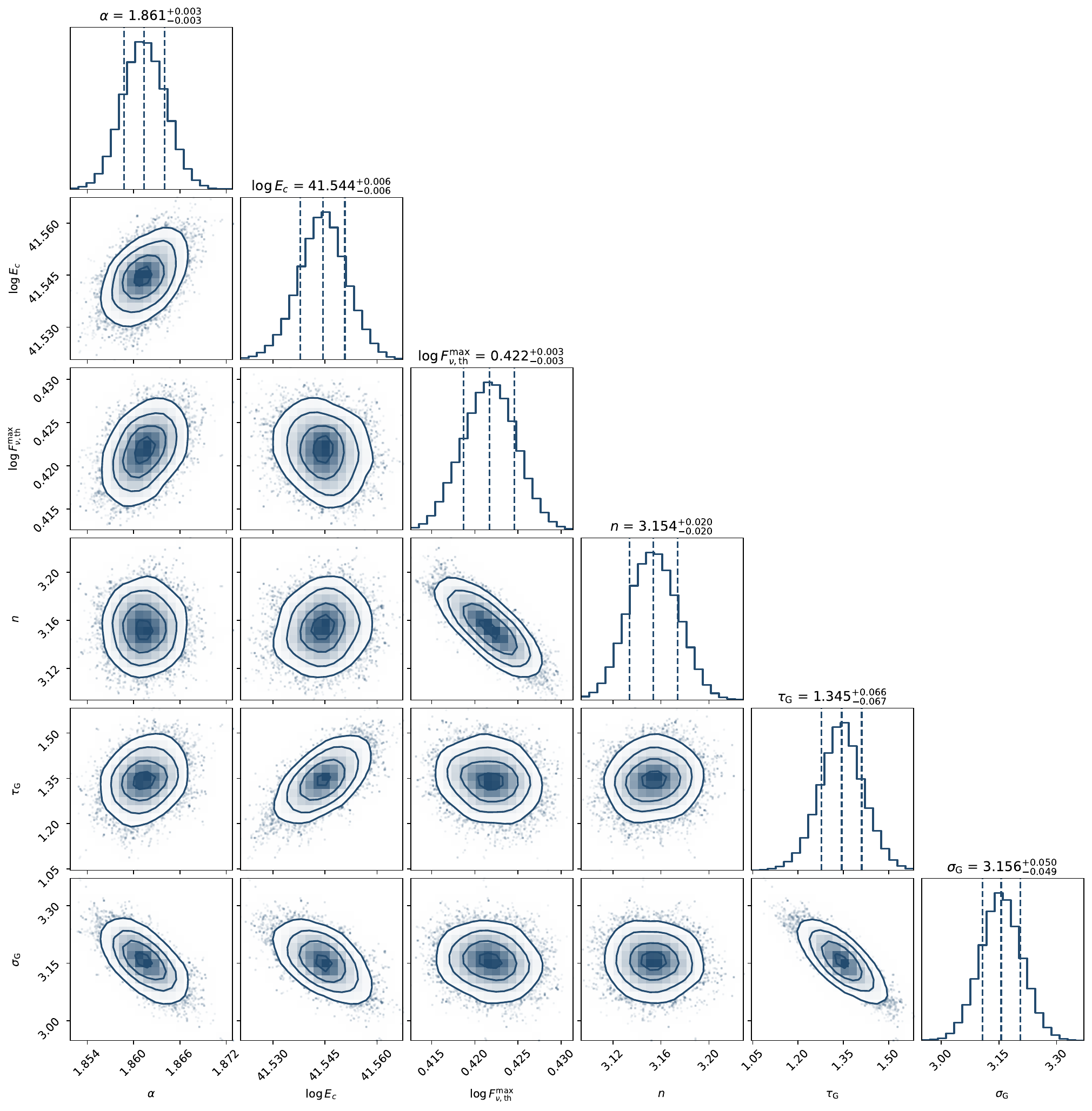}{0.48\textwidth}{}
}
\gridline{
    \fig{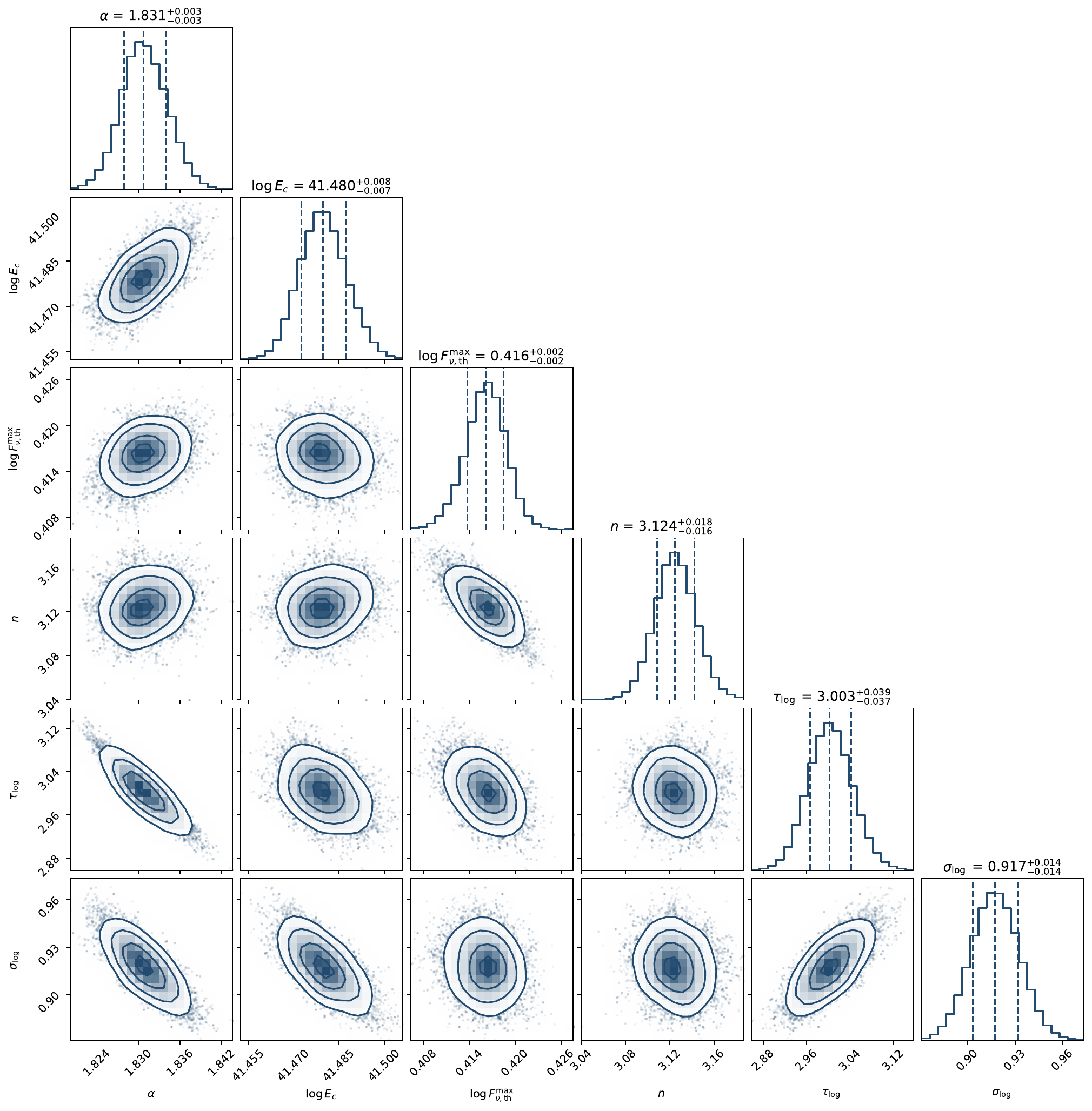}{0.48\textwidth}{}
    \fig{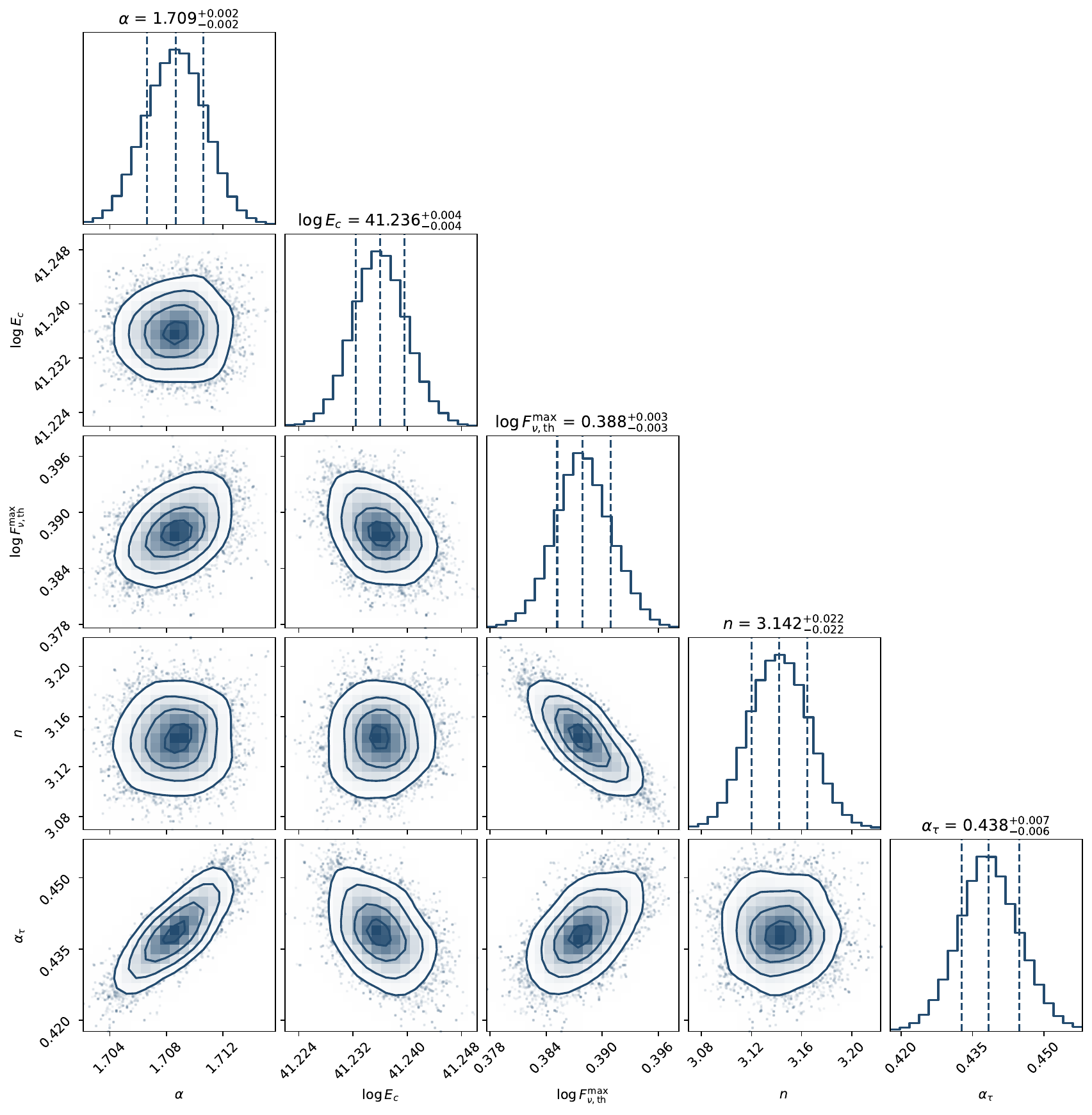}{0.48\textwidth}{}
}
\caption{Corner plots of the four phenomenological models considered in this work. The panels correspond to the SFH model (top left), Gaussian delay model (top right), log-normal delay model (bottom left), and power-law delay model (bottom right), respectively.}
\label{fig:phenomenological_corner}
\end{figure*}

The prior ranges adopted for all free parameters are summarized in Table~\ref{table1}. The best fit parameters are summarized in Tables~\ref{tab2}, \ref{tab:common_params}, and \ref{tab:model_specific_params}. For model comparison, we use the Bayesian information criterion (BIC),
\begin{equation}
\mathrm{BIC}=-2\ln \mathcal{L}_{\max}+k\ln N,
\end{equation}
where $\mathcal{L}_{\max}$ is the maximum value of the likelihood, $k$ is the number of free parameters in a given model, and $N$ denotes the number of data points entering the likelihood calculation. Since the second term penalizes model complexity, a smaller BIC indicates a more favorable balance between goodness of fit and the number of fitted parameters. Among the phenomenological models, the pure SFH model provides the poorest description of the Catalog 2 Gold Sample, with $\mathrm{BIC}=21146.27$, indicating that the one-off FRB population is not well described by a redshift distribution that directly traces the SFH. All three delayed phenomenological models perform substantially better. The Gaussian delay model yields the lowest BIC, $\mathrm{BIC}=5008.74$, followed by the log-normal model ($6789.07$) and the power-law delay model ($9018.75$). This behavior is also illustrated in Figure~\ref{fig:mixture_overlay}, where the model predictions are compared directly with the Gold Sample distributions in $\log E$, redshift, and $\log F_{\nu}$. The SFH-tracing model shows the largest mismatch with the observed distributions, whereas the delayed models provide a visibly improved description, especially in the redshift distribution. The mixture model, included in the same comparison, gives the best overall agreement with the observed one-dimensional distributions, consistent with its substantially lower BIC. The common parameters remain broadly stable across these fits, with $\alpha \simeq 1.7$--$1.9$, $\log E_c \simeq 41.2$--$41.5$, $\log F_{\nu,\mathrm{th}}^{\max} \simeq 0.39$--$0.42$, and $n \simeq 3.1$--$3.2$, suggesting that the model ranking is driven mainly by the assumed redshift evolution. The delayed models favor non-negligible delay scales. For the Gaussian model, the fit gives $\tau_{\rm G}=1.34^{+0.07}_{-0.07}~\mathrm{Gyr}$ and $\sigma_{\rm G}=3.16^{+0.05}_{-0.05}~\mathrm{Gyr}$. Since the distribution is truncated at $\tau=0$, the location parameter $\tau_{\rm G}$ is not itself the mean delay. Instead, the mean delay should be computed from the truncated distribution as $\bar{\tau} = \int_0^\infty \tau f(\tau)\,d\tau \big/ \int_0^\infty f(\tau)\,d\tau$, which gives $\bar{\tau}=3.073~\mathrm{Gyr}$ for the best fit Gaussian model. For the log-normal model, the fit yields $\tau_{\log}=3.00^{+0.04}_{-0.04}~\mathrm{Gyr}$ and $\sigma_{\log}=0.92^{+0.01}_{-0.01}$, corresponding to a mean delay of $\bar{\tau}=4.567~\mathrm{Gyr}$. The power-law delay model instead prefers a shallow delay-time slope, $\alpha_{\tau}=0.438^{+0.007}_{-0.006}$. Taken together, these results favor a delayed population over one that directly follows the SFH. The posterior for the four phenomenological models are shown in Figure~\ref{fig:phenomenological_corner}. These corner plots provide a visual check of the marginalized constraints on the common population parameters and the delay-time parameters used in the SFH, Gaussian, log-normal, and power-law models.

\begin{figure*}
\centering
\includegraphics[width=0.95\textwidth]{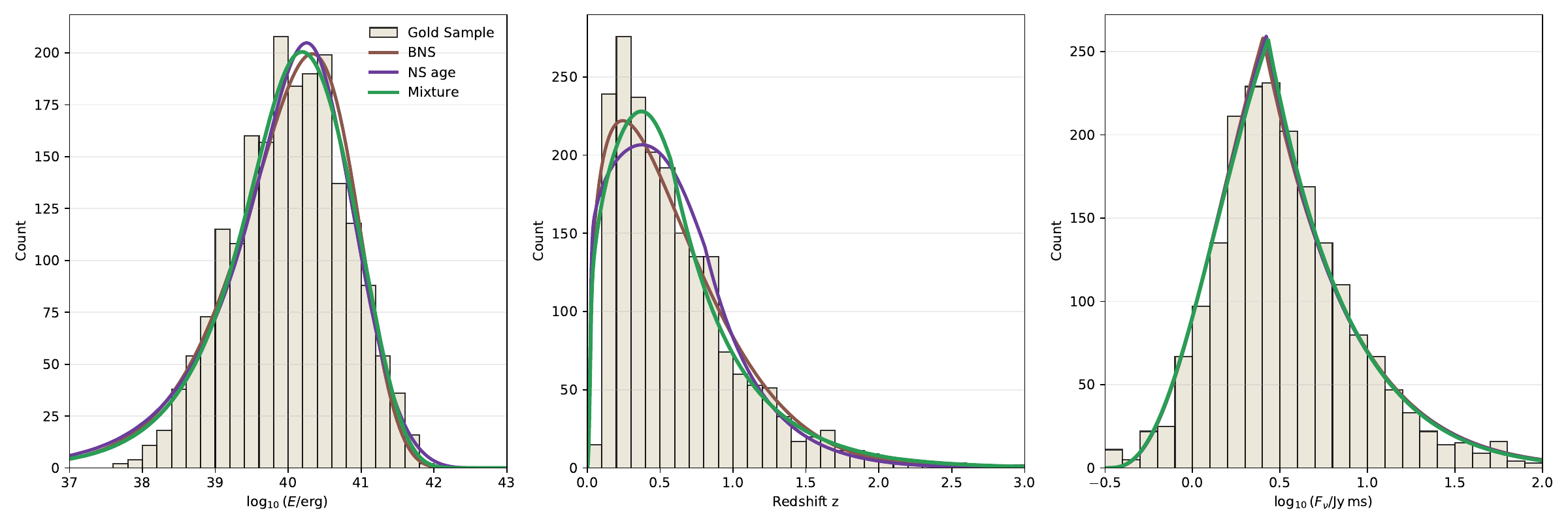}
\caption{
Comparison between the Gold Sample distributions and the physically motivated models. 
The three panels show the distributions of $\log E$, redshift $z$, and $\log F_{\nu}$, respectively. 
The pure BNS-related model, the pure neutron-star age-window model, and the mixture model are shown together. 
The mixture model provides the best simultaneous description of the observed distributions, consistent with its lowest BIC.
}
\label{fig:physical_overlay}
\end{figure*}

\begin{figure*}[htbp]
\centering
\gridline{
    \fig{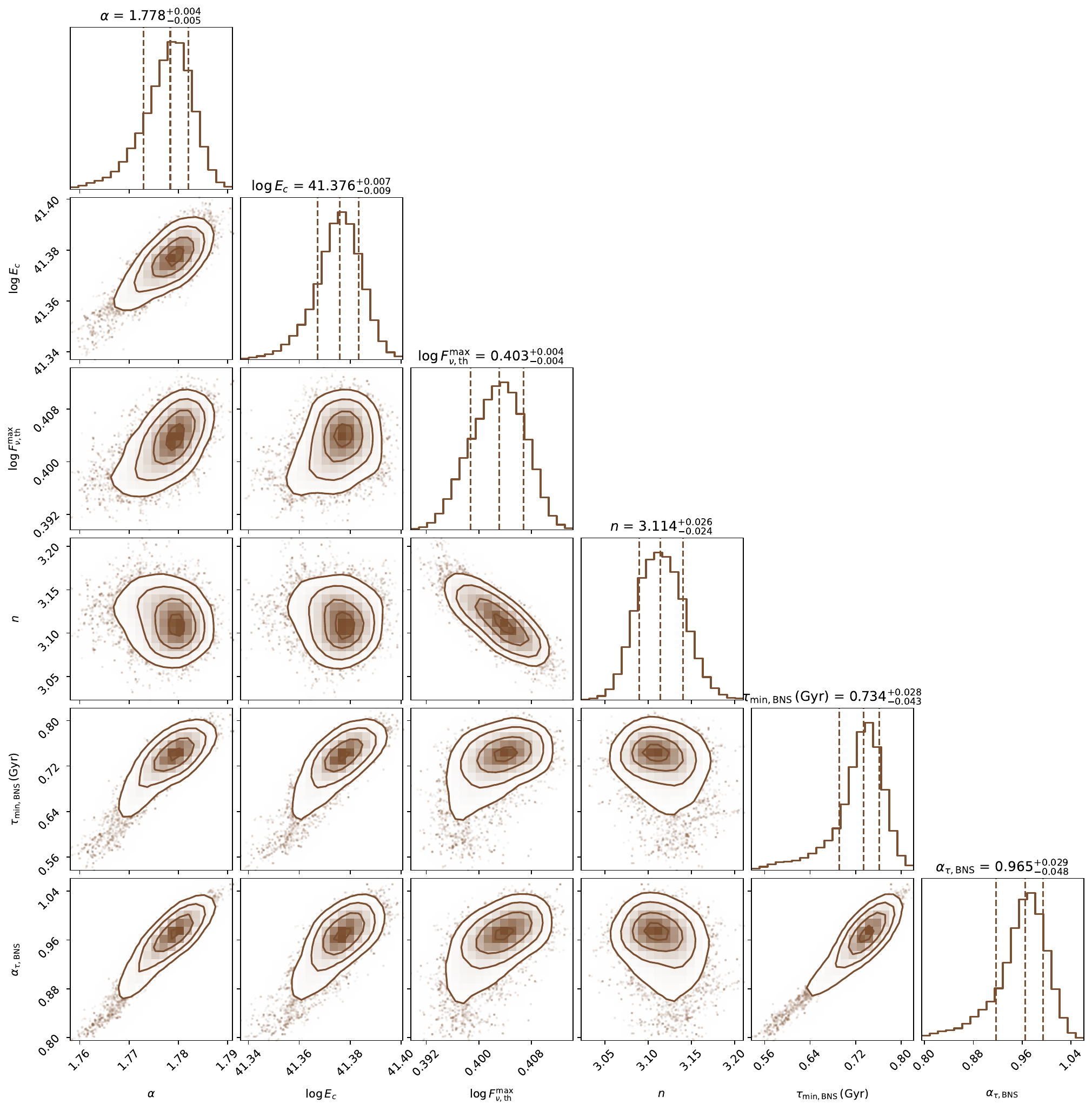}{0.48\textwidth}{}
    \fig{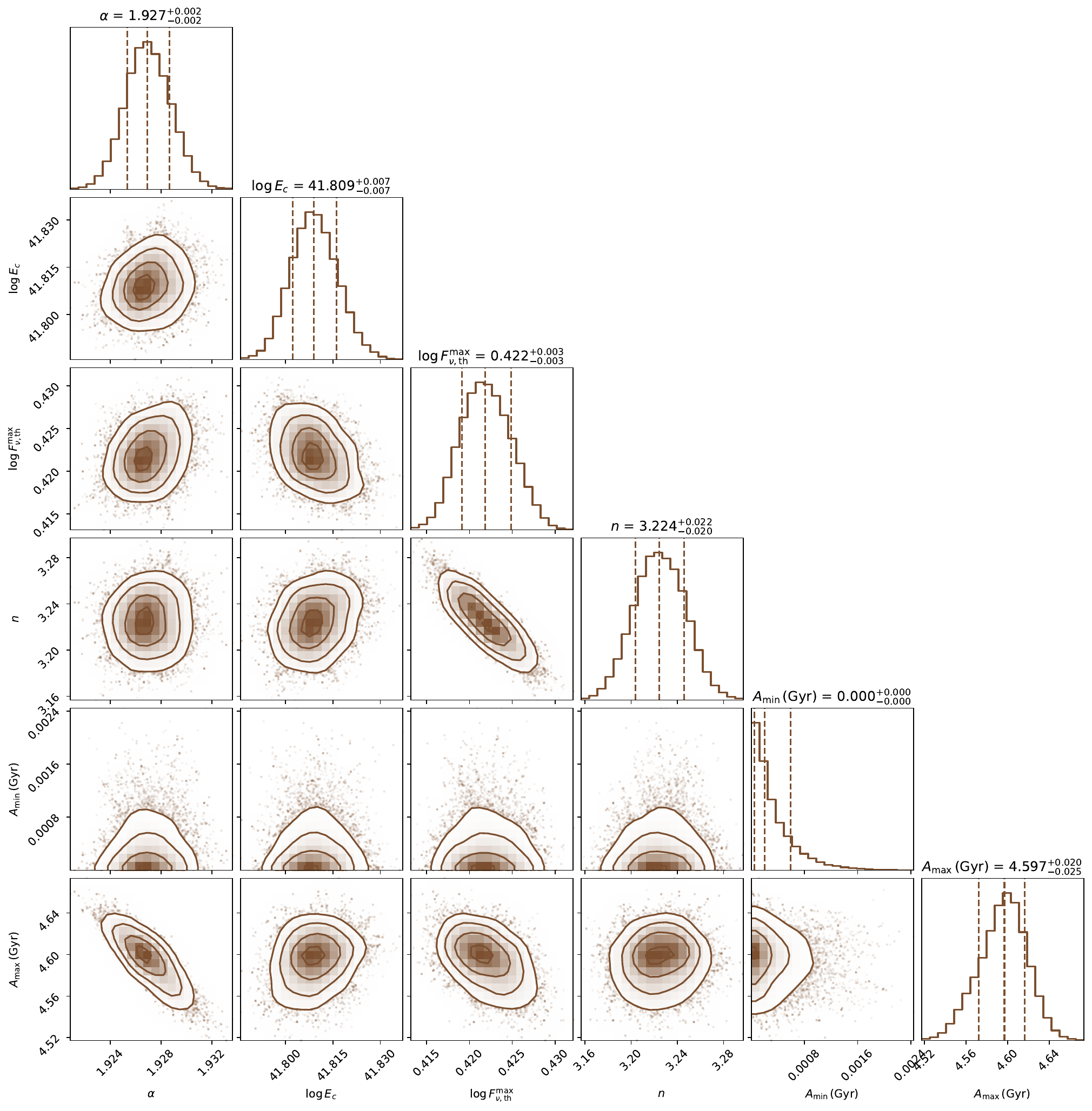}{0.48\textwidth}{}
}
\gridline{\fig{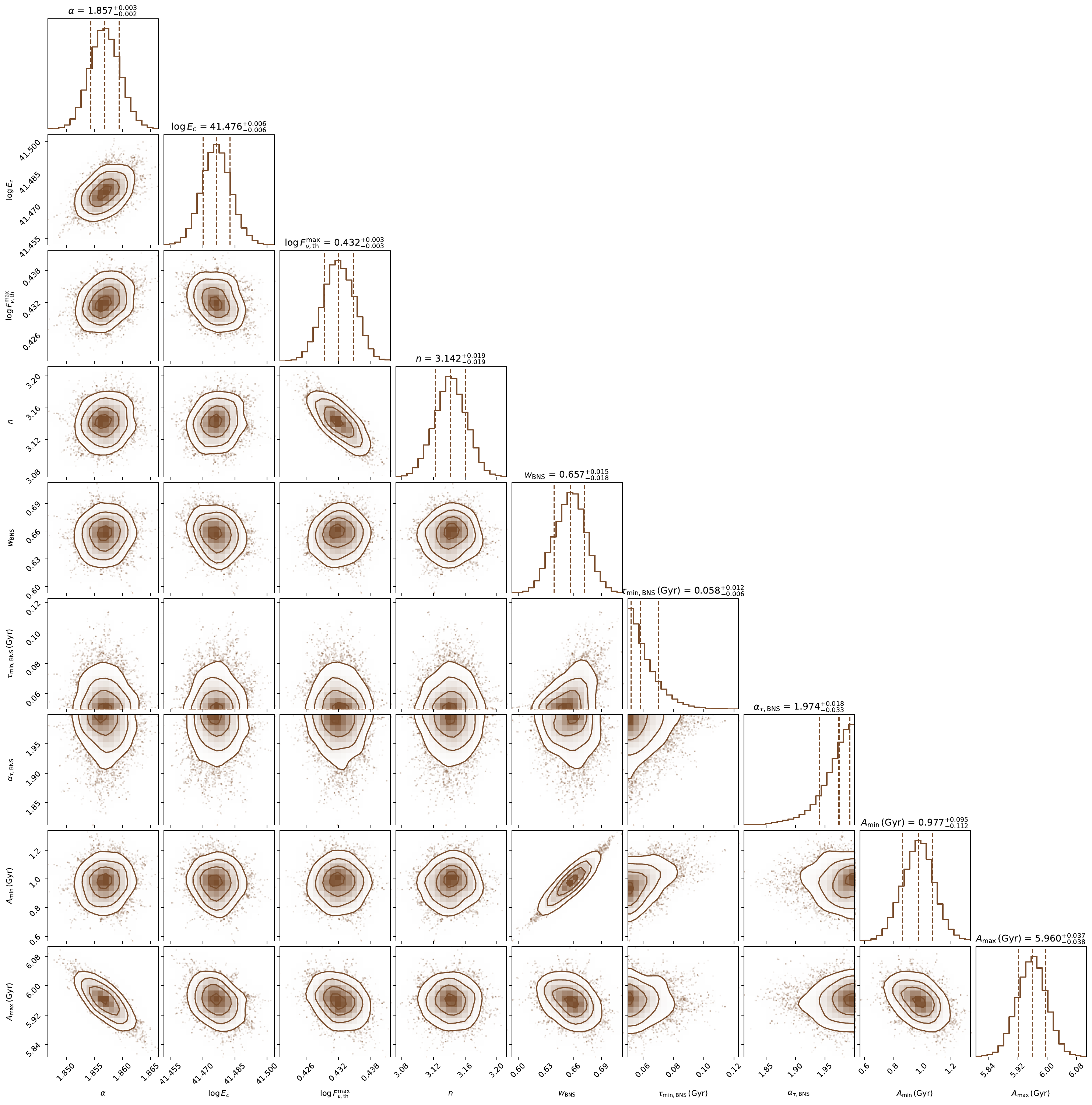}{0.48\textwidth}{}}
\caption{Corner plots of the three physically motivated models considered in this work. The top-left, top-right, and bottom panels correspond to the BNS-related model, the neutron star age-window model, and the mixture model, respectively.}
\label{fig:physical_corner}
\end{figure*}

The physically motivated models show an even stronger preference for a mixed origin. The mixture model gives the lowest BIC of all models considered, $\mathrm{BIC}=3653.80$, improving over the pure BNS model by $4264.28$, over the pure NS age-window model by $4889.77$, and over the best phenomenological model by $1354.94$. The corresponding distribution-level comparison is shown in Figure~\ref{fig:physical_overlay}. The pure BNS, pure NS age-window, and mixture models are overlaid on the Gold Sample distributions in $\log E$, redshift, and $\log F_{\nu}$. Although the two single-component physical models can reproduce part of the observed distribution, the mixture model provides the most consistent simultaneous match to the energy, redshift, and fluence distributions.  For the pure BNS model, the fit prefers $\tau_{\rm min,BNS}=0.73^{+0.03}_{-0.04}~\mathrm{Gyr}$ and $\alpha_{\tau,\rm BNS}=0.96^{+0.03}_{-0.05}$. In the mixture model, however, the BNS-related component shifts to a much shorter minimum delay and a steeper delay-time distribution, with $\tau_{\rm min,BNS}=0.058^{+0.012}_{-0.006}~\mathrm{Gyr}$ and $\alpha_{\tau,\rm BNS}=1.974^{+0.018}_{-0.033}$. The NS age-window component also changes between the single-component and mixture fits. In the pure age-window model, the lower bound is driven to the prior edge, $A_{\min}\approx 0$, while the upper bound is constrained to $A_{\max}=4.597^{+0.020}_{-0.025}~\mathrm{Gyr}$. In the mixture model, the preferred window shifts to $A_{\min}=0.98^{+0.09}_{-0.11}~\mathrm{Gyr}$ and $A_{\max}=5.960^{+0.037}_{-0.038}~\mathrm{Gyr}$. The inferred mixture weight is $w_{\rm BNS}=0.657^{+0.015}_{-0.018}$. The corresponding posterior distributions for the physically motivated models are shown in Figure~\ref{fig:physical_corner}. We obtain $\bar{\tau}=4.7256^{+0.0683}_{-0.0500}~\mathrm{Gyr}$ for the pure BNS model, $\bar{\tau}=2.2986^{+0.0098}_{-0.0123}~\mathrm{Gyr}$ for the pure NS age-window model, and $\bar{\tau}=1.4264^{+0.0324}_{-0.0352}~\mathrm{Gyr}$ for the mixture model. These values show that the pure BNS model corresponds to the longest characteristic delay, while the mixture model implies the shortest effective delay among the physically motivated cases considered here. The mean delay time inferred for our mixture model is broadly consistent with recent FRB population studies that suggest characteristic delay times of order $\sim 1~\mathrm{Gyr}$ \citep{2024A&A...690A.377W, 2025ApJ...986..100G, 2025A&A...698A.127M}.

Overall, the fits do not support a one-off FRB population that directly traces the SFH. Instead, they favor delayed redshift evolution, with the physically motivated models explored here pointing to a two components interpretation that provides the best statistical description of the sample, consistent with the mixed origin scenario discussed by \citet{2025ApJ...986..100G}, who used CHIME Catalog 1. This result should not be over-interpreted as direct proof of multiple progenitor channels, since it remains conditional on the adopted $\mathrm{DM}_E$--$z$ relation, selection treatment, and specific parameterizations of $R(z)$. Nevertheless, the substantial BIC improvement suggests that a hybrid physical scenario is a plausible interpretation of the present data.

\section{Conclusion and Discussion}
\label{sec:conclusion}

In this work, we investigated the evolution of one-off FRBs using the CHIME/FRB Catalog 2 Gold Sample and an updated $\mathrm{DM}_{\rm E}$--$z$ relation. Because most CHIME FRBs do not have confirmed host galaxy redshifts, we inferred the redshift information statistically from the extragalactic dispersion measure and compared the resulting distributions with several models for the intrinsic FRB event rate density. We modeled the observed fluence, energy, and redshift distributions simultaneously, including an empirical detection efficiency function and a power-law burst energy distribution with an exponential cutoff. This framework enables us to examine whether the one-off FRB population directly traces the SFH, is better represented by a delayed formation channel, or instead favors a physically motivated models.

Our main result is that the one-off FRB population is not well described by a rate density that directly traces the SFH. The pure SFH model gives the poorest fit among all models considered, with 
$\mathrm{BIC}=21146.27$. In contrast all delayed phenomenological models provide substantially better descriptions of the data. Among them, the Gaussian delay model gives the lowest BIC, $\mathrm{BIC}=5008.74$, followed by the log-normal and power-law delay models. The preferred parameters of the burst energy distribution and detection function remain broadly similar across these delayed models, suggesting that the improvement is driven mainly by the redshift evolution rather than by changes in the assumed energy function or selection prescription. This provides evidence that the apparent redshift distribution of one-off FRBs favors a delayed evolutionary history instead of a prompt SFH-tracing origin.

We further tested three physically motivated models: a BNS-related model, a neutron star age-window model, and a mixture of the two. The BNS-related model assumes that FRBs are associated with binary neutron star mergers, occurring either promptly at merger or shortly thereafter. The neutron star age-window model instead assumes that FRBs are produced only during a finite active phase after neutron star formation. The mixture model gives the best fit among all models examined in this work, with $\mathrm{BIC}=3653.80$. This represents a substantial improvement over the pure BNS model, the pure neutron-star age-window model, and the best phenomenological model. In the mixture fit, the BNS-related component favors a short minimum delay, $\tau_{\rm min,BNS}=0.058^{+0.012}_{-0.006}~\mathrm{Gyr}$, and a steep delay time distribution, $\alpha_{\tau,\rm BNS}=1.974^{+0.018}_{-0.033}$. The neutron-star age-window component is constrained to $A_{\min}=0.98^{+0.09}_{-0.11}~\mathrm{Gyr}$ and $A_{\max}=5.960^{+0.037}_{-0.038}~\mathrm{Gyr}$. The inferred mixture weight is $w_{\rm BNS}=0.657^{+0.015}_{-0.018}$, indicating that the redshift distribution is better reproduced when a dominant BNS-related component is combined with a subdominant delayed neutron-star age-window component. The steep delay time distribution inferred for the BNS-related component in the mixture model does not necessarily imply that one-off FRBs preferentially arise from short delay mergers. Instead, it may reflect the contribution of an unmodeled underlying channel that is incorrectly absorbed into the binary neutron star component.

The physical implication of this result is that one-off FRBs may have multi-channel progenitor. A BNS-related contribution is consistent with scenarios in which FRBs are produced either during compact object merger activity or by young magnetars formed in merger remnants. The neutron star age-window component assumes that a neutron star must evolve for at least $A_{\rm min}$ Gyrs before it can produce a one-off FRB. The effective mean delay time inferred for the mixture model is $\bar{\tau}=1.426^{+0.032}_{-0.035}~\mathrm{Gyr}$. This results is consistent with some analysis that characteristic delay times of one-off FRBs is of order $\sim 1~\mathrm{Gyr}$ \citep{2024A&A...690A.377W, 2025ApJ...986..100G, 2025A&A...698A.127M}. We caution that the mixture weight $w_{\rm BNS}$ should not be interpreted as a direct astrophysical branching fraction. Since the two components are individually normalized before being combined, $w_{\rm BNS}$ instead quantifies the relative contribution of two redshift-distribution shapes within the adopted model space, rather than the intrinsic fraction of all FRBs produced by BNS-related systems. Likewise, while the BIC comparison identifies the parameterization preferred by the present data, it does not by itself establish the existence of a specific progenitor channel. 

Therefore, our results suggest a picture in which the observed one-off FRB population is shaped by delayed progenitor evolution, with the best current description being a hybrid channel rather than a purely prompt or purely single delay population.

Despite these limitations, the strong statistical preference for delayed and mixed models has important implications for FRB progenitor studies. It suggests that the redshift distribution of one-off FRBs contains information about progenitor evolution beyond that encoded in the SFH alone. Future progress may also be driven by the next generation of gravitational wave observations. Third generation detectors, such as the Einstein Telescope and Cosmic Explorer \citep{2010CQGra..27s4002P, 2021arXiv210909882E, 2023JCAP...07..068B}, are expected to greatly expand the detectable sample of BNS mergers and may improve the prospects for constraining gravitational wave signals from neutron star oscillations. These observations would provide an independent probe of compact object populations and could help test whether delayed FRB components are associated with BNS-related channels, long-lived neutron star remnants, or other evolved neutron star populations. Within the present framework, however, the CHIME/FRB Catalog 2 Gold Sample favors a delayed, hybrid origin for one-off FRBs.

\begin{acknowledgments}
We thanks Ying-Ze Shan for helpful discussion.
\end{acknowledgments}

\begin{table}[htbp]
\centering
\caption{Prior ranges of the free parameters used in different models. Here, $\mathcal{U}(a,b)$ denotes a uniform prior over the interval $(a,b)$. The parameters $A_{\min}$, $A_{\max}$, $\tau_{\log/G}$, $\sigma_{\log/G}$ and $\tau_{\rm min,BNS}$ are in units of Gyr.}
\label{table1}
\begin{tabular}{lll}
\hline
\hline
Parameter & Prior & Applicable model \\
\hline
$\alpha$ & $\mathcal{U}(0.1,\,5)$ & All models \\
$\log E_c$ & $\mathcal{U}(37,\,44)$ & All models \\
$\log F^{\max}_{\nu,\rm th}$ & $\mathcal{U}(0.01,\,2)$ & All models \\
$n$ & $\mathcal{U}(0.1,\,8)$ & All models \\
\hline
$\alpha_{\tau}$ & $\mathcal{U}(0.1,\,5)$ & Power-law DTD \\
\hline
$\tau_{\log}$ & $\mathcal{U}(0.1,\,5)$ & Log-normal DTD \\
$\sigma_{\log}$ & $\mathcal{U}(0.1,\,5)$ & Log-normal DTD \\
\hline
$\tau_{\rm G}$ & $\mathcal{U}(0.1,\,5)$ & Gaussian DTD \\
$\sigma_{\rm G}$ & $\mathcal{U}(0.1,\,5)$ & Gaussian DTD \\
\hline
$A_{\min}$ & $\mathcal{U}(0,\,13.8)$ & NS age-window / Mixture \\
$A_{\max}$ & $\mathcal{U}(0.001,\,13.8)$ & NS age-window / Mixture \\
$\tau_{\rm min,BNS}$ & $\mathcal{U}(0,\,5)$ & BNS / Mixture \\
$\alpha_{\tau,\rm BNS}$ & $\mathcal{U}(0.8,\,2)$ & BNS / Mixture \\
$w_{\rm BNS}$ & $\mathcal{U}(0,\,1)$ & Mixture\\
\hline
\end{tabular}
\end{table}

\begin{table*}[htbp]
\centering
\caption{Best-fit parameters and BIC values for the phenomenological models fitted to the catalog2 Gold Sample using
the \citet{2026arXiv260318487C} DM$_E$-$z$ relation.}
\label{tab2}
\begin{tabular}{lccccccc}
\toprule
Model & $\alpha$ & $\log E_c/\mathrm{erg}$ & $\log F_{\nu,\mathrm{th}}^{\max}$ & $n$ & \multicolumn{2}{c}{Model
  Parameter} & $\mathrm{BIC}$ \\
\hline
SFH
& $1.914^{+0.001}_{-0.001}$
& $40.925^{+0.002}_{-0.002}$
& $0.614^{+0.002}_{-0.002}$
& $3.23^{+0.01}_{-0.01}$
& N/A
& N/A
& 21146.27 \\

Gaussian
& $1.861^{+0.003}_{-0.003}$
& $41.544^{+0.006}_{-0.006}$
& $0.422^{+0.003}_{-0.003}$
& $3.15^{+0.02}_{-0.02}$
& $\tau_{\rm G} = 1.34^{+0.07}_{-0.07}$
& $\sigma_{\rm G} = 3.16^{+0.05}_{-0.05}$
& 5008.74 \\

Log-normal
& $1.831^{+0.003}_{-0.003}$
& $41.480^{+0.008}_{-0.007}$
& $0.416^{+0.002}_{-0.002}$
& $3.12^{+0.02}_{-0.02}$
& $\tau_{\log} = 3.00^{+0.04}_{-0.04}$
& $\sigma_{\log} = 0.92^{+0.01}_{-0.01}$
& 6789.07 \\

Power-law
& $1.709^{+0.002}_{-0.002}$
& $41.236^{+0.004}_{-0.004}$
& $0.388^{+0.003}_{-0.003}$
& $3.14^{+0.02}_{-0.02}$
& $\alpha_{\tau} = 0.438^{+0.007}_{-0.006}$
& N/A
& 9018.75 \\

\hline
\end{tabular}
\end{table*}

\begin{table*}[htbp]
\centering
\footnotesize
\setlength{\tabcolsep}{6pt}
\renewcommand{\arraystretch}{1.1}
\caption{Best-fit common parameters for the BNS, NS age-window, and mixture models.}
\label{tab:common_params}
\begin{tabular}{lcccc}
\toprule
Model & $\alpha$ & $\log E_c/\mathrm{erg}$ & $\log F_{\nu,\mathrm{th}}^{\max}$ & $n$ \\
\midrule
BNS
& $1.778^{+0.004}_{-0.005}$
& $41.376^{+0.007}_{-0.009}$
& $0.403^{+0.004}_{-0.004}$
& $3.11^{+0.03}_{-0.02}$ \\

NS age-window
& $1.927^{+0.002}_{-0.002}$
& $41.809^{+0.007}_{-0.007}$
& $0.422^{+0.003}_{-0.003}$
& $3.22^{+0.02}_{-0.02}$ \\

Mixture
& $1.857^{+0.003}_{-0.002}$
& $41.476^{+0.006}_{-0.006}$
& $0.432^{+0.003}_{-0.003}$
& $3.14^{+0.02}_{-0.02}$ \\
\bottomrule
\end{tabular}
\end{table*}

\begin{table*}[htbp]
\centering
\footnotesize
\setlength{\tabcolsep}{6pt}
\renewcommand{\arraystretch}{1.1}
\caption{Best-fit model specific parameters and $\mathrm{BIC}$ values for the BNS, NS age-window, and mixture models. The parameters $\tau_{\rm min,BNS}$, $A_{\min}$, and $A_{\max}$ are in units of Gyr.}
\label{tab:model_specific_params}
\begin{tabular}{lcccccc}
\toprule
Model & $\tau_{\rm min,BNS}$ & $\alpha_{\tau,\rm BNS}$ & $A_{\min}$ & $A_{\max}$ & $w_{\rm BNS}$ & $\mathrm{BIC}$ \\
\midrule
BNS
& $0.73^{+0.03}_{-0.04}$
& $0.96^{+0.03}_{-0.05}$
& N/A
& N/A
& N/A
& 7918.08 \\

NS age-window
& N/A
& N/A
& $0.00020^{+0.00039}_{-0.00015}$
& $4.597^{+0.020}_{-0.025}$
& N/A
& 8543.57 \\

Mixture
& $0.058^{+0.012}_{-0.006}$
& $1.974^{+0.018}_{-0.033}$
& $0.98^{+0.09}_{-0.11}$
& $5.960^{+0.037}_{-0.038}$
& $0.657^{+0.015}_{-0.018}$
& 3653.80 \\
\bottomrule
\end{tabular}
\end{table*}

% In each panel, the diagonal subplots show the one-dimensional marginalized posterior distributions of the model parameters, while the off-diagonal subplots show the corresponding two-dimensional joint posterior distributions. These plots illustrate the parameter constraints and degeneracies for the phenomenological descriptions of the one-off FRB population.

%In each panel, the diagonal subplots show the one-dimensional marginalized posterior distributions of the model parameters, while the off-diagonal subplots show the corresponding two-dimensional joint posterior distributions. These plots illustrate the parameter constraints and degeneracies in the physically motivated descriptions of the one-off FRB population.}

\end{document}